\documentclass[10pt,a4paper,twoside,bibliography=totocnumbered]{article}

\usepackage{xcolor}

\usepackage{graphicx}

\usepackage{amsmath}
\usepackage{amssymb}

\usepackage{geometry}

\usepackage{pdflscape}

\begin{document}

\title{{\huge \textcolor{black}{Random Inverse Problems with Structural and Probabilistic Ambiguities}\vspace*{-0.4cm}\\ \begin{minipage}{5cm}\centering \textcolor{black}{---}\vspace*{-0.4cm}
\end{minipage}\\}{\Large \textcolor{black}{Original Research Article}}\vspace*{0.5cm}}


\author{Wolfgang Hoegele$^{1}$\vspace*{0.4cm}\\
{\normalsize $^{1}$ Munich University of Applied Sciences HM}\\{\normalsize Department of Computer Science and Mathematics}\\{\normalsize Lothstraße 64, 80335 München, Germany}\vspace*{0.4cm}\\
{\normalsize corresponding mail: \texttt{wolfgang.hoegele@hm.edu}}\vspace*{0.4cm}\\
ORCID: 0000-0002-5303-9334 \vspace*{0.4cm}}

\date{\today}

\maketitle
\thispagestyle{empty}

\section*{Abstract}

In this paper, we investigate a computational class of \textit{random inverse problems} that incorporates model uncertainties through random variable parameters nonlinearly in the forward model as well as additive observational uncertainty. Random inverse problems with nonlinear parameter dependencies may arise in engineering, geophysics, image processing or uncertainty quantification. We study a perspective on structural ambiguities due to the non-injectivity of the forward model together with probabilistic ambiguities by assigning mixture model densities with separate components to the parameters, which leads to an in general complex forward model, observation model and posterior. As a result, the mixture-model parameters in the forward model can be viewed as simultaneously describing aspects of both the nonlinear ambiguity and the uncertainty of the inverse problem. The presented solution algorithms are based on Bayesian inversion and Monte Carlo computation for three observation scenarios leading to posterior densities for the input for given output samples or an observed output density. We apply the derived algorithms to analytic 1D and 2D quadratic models, to an epidemiological inverse parameter estimation and to a heat equation for inverse source localization. We numerically demonstrate in which observation scenarios the proposed algorithm can resolve probabilistic ambiguities in the solution and in which they cannot and show the interaction patterns between these two types of ambiguities. The results suggest that the proposed perspective is useful in making residual structural ambiguities visible in highly ambiguous inverse problems, including cases with a finite and an infinite number of solutions.\medskip

\textbf{Keywords:} random inverse problems, stochastic modeling, random forward problem, posterior density, non-injective function\medskip

\newpage
\thispagestyle{empty}
\tableofcontents

\section*{About the Author}

Dr. Högele is Professor of Applied Mathematics and Computational Science at the Department of Computer Science and Mathematics at the Munich University of Applied Sciences (HM), Germany. His research interests are in general mathematical modeling and, more specifically, stochastic modeling, simulation, and analysis of complex systems in applied mathematics.\medskip

\thispagestyle{empty}

\newpage

\section{Introduction}

Inverse problems are an important class of problems in applied mathematics and are defined by an input space, an output space, and a forward model which maps elements of the input to the output space. Given an output observation the goal is to find the corresponding inputs that best explain the output. The main challenge is that the inverse problem is often ill-posed, which typically means that the forward model is in theory or in numerical practice not readily invertible \cite{Calvetti 2018}. Incorporating uncertainties in the inverse problem further increases the problem difficulty and the solution is typically described by an input probability density. A common approach for such uncertainty-driven inverse problems is regularization in the framework of \textit{Bayesian inverse problems} \cite{Calvetti 2018, Stuart 2010, Dashti 2017}. Many common formulations make assumptions that we want to relax in this publication: 

First assumption: the forward model itself is known deterministically (only observational noise introduces uncertainty) and it is smooth to some degree, e.g. as a solution of differential equations, or even mostly linear in studies, e.g. cp. to \cite{Kaipio 2007,Sanz-Alonso 2025, Saibaba 2019}. Quite common is the treatment of nonlinear forward models by discretized linearizations, e.g. \cite{Saibaba 2019}. We allow directly for nonlinear and only piecewise continuous forward model functions which is uncommon in this generality. 

Second assumption: stochastic uncertainty is typically introduced only by additive observational noise at the output which is a comparably simple way of introducing uncertainty, e.g. see \cite{Kaipio 2007,Sanz-Alonso 2025, Saibaba 2019}. In this work, we allow the in general nonlinear incorporation of uncertainty in the forward model as part of a challenging uncertainty propagation, i.e. we regard the forward model itself as stochastic. This means, it depends on random variable parameters which have possibly different realizations for every forward model evaluation which we call a \textit{random forward / inverse problem}. This highlights the close relation to nonlinear \textit{random equations} and although not a broad standard term it is appropriate and used before accordingly \cite{Zhang 2025}.

Third assumption: many widely used formulations use comparably simple densities for observational uncertainty and more complicated non-Gaussian and non-uniform distributions are rarely presented, e.g. see \cite{Kaipio 2007,Sanz-Alonso 2025,Saibaba 2019}. We study complicated probability densities by mixture models inside the forward model, which represent a flexible class of multi-modal densities. In total, this relates to inverse problems with Monte Carlo simulators \cite{Cranmer 2026} although we utilize mixture models in order to capture the forward model complexity instead of a black box simulator.
 
In particular, we focus on non-injective ill-posed inverse problems, i.e. different inputs may lead to the same output of the forward model which introduces ambiguity in the solution of the inverse problem, e.g. see \cite{Sun 2022}. Further, we introduce probabilistic ambiguities by utilizing random parameter densities inside the forward model that contain mixture models with separated components or distinct modes. Such probabilistic ambiguities are not part of classical Bayesian inversion presentations and a new addition in this work. Investigating these two types of ambiguities simultaneously is part of the special setup and motivates the general formulation in this study. We further differentiate between three observational scenarios (i)-(iii) which illustrates the flexibility of the presented methodology and provides deeper practical insights on how structural and probabilistic ambiguities interact in the solution of inverse problems. It should be noted that discussing estimators for this type of problem with ambiguities is typically inappropriate since the solution densities are by definition multi-modal (often with separated similar peaks as will be presented in the numerical examples) and defining maxima (such as for Maximum a Posteriori) or averages (such as for Mean Squared Error estimators), e.g. see \cite{Chada 2026}, for such densities does not suite this problem class. In consequence, we will focus on full density presentations.

These described extensions are the new conceptual reframing of the random inverse problem and provide an extended perspective and argumentation framework for ambiguity treatment compared to standard inverse problems. For the derivation of the actual solution algorithms Bayesian likelihood / posterior formulations and Monte Carlo techniques are utilized and they are in line with recently published derivations which have a broad application appeal, e.g. such as for random equations \cite{Hoegele 2026}, stochastic dynamical systems \cite{Hoegele 2026 dyn1}, model fitting \cite{Hoegele 2025 stat} and computer vision \cite{Hoegele 2024 imag}. These surrounding works provide an instructive context for the direct derivations, although they are not directly concerned with ill-posed inverse problems. Since the presented methodology has generality of the stochastic models in focus including accessible general computational algorithms, i.e. allowing for many possible applications, straightforward convergence or consistency theorems are not part of this exploratory presentation, which are typically only possible for selected model and density classes.

The outline of the paper is as follows: In the methods section, the definition, general posterior derivation and observation scenarios as well as algorithmic implementations are presented. In the results section, numerical simulation examples for analytic as well as computational models are presented which help to illustrate this methodology.

\section{Methods}

\subsection{Random Inverse Problem Formulation and Observation Scenarios}
\label{sec:GenProbForm}

\textit{Random inverse problems} contain random variables (in general nonlinearly) in the forward model as well as additive observational uncertainties following the general equation
\begin{align}
\boldsymbol{y} = \boldsymbol{M}(\boldsymbol{x};\boldsymbol{A}) + \boldsymbol{B}\;,
\end{align}
with $\boldsymbol{y}\in\mathbb{R}^R$ the given output of the system, $\boldsymbol{x}\in\mathbb{R}^n$ the input we want to reconstruct, $\boldsymbol{A}$ a $K$-dimensional random variable vector and $\boldsymbol{B}$ an $R$-dimensional observational uncertainty/noise random variable vector. The system function $\boldsymbol{M}: \mathbb{R}^n\times \mathbb{R}^K\rightarrow \mathbb{R}^R$ is at least piecewise continuous and in general nonlinear with respect to $\boldsymbol{x}$ and $\boldsymbol{A}$. We deliberately distinguish between additive observational uncertainties in $\boldsymbol{B}$ and inherent model randomness in $\boldsymbol{A}$ since we regard this as different sources of uncertainties. The goal of the inverse problem is to find $\boldsymbol{x}$ which explains $\boldsymbol{y}$ under $\boldsymbol{M}$ considering $\boldsymbol{A}$ and $\boldsymbol{B}$. In classical Bayesian inverse problems this relation is described with a deterministic system function $\boldsymbol{M}$, a random variable input $\boldsymbol{x}$ and only additive random noise, i.e. $\boldsymbol{y} = \boldsymbol{M}(\boldsymbol{x}) + \boldsymbol{B}$ (in many publications even the strong limitation to $\boldsymbol{M}(\boldsymbol{x}) = M\, \boldsymbol{x}$ with a $M\in\mathbb{R}^{R\times n}$ matrix is used). Another difference is that we consider $\boldsymbol{x}$ not a random variable in the forward model but it becomes a random variable when solving the inverse problem by Bayesian methods. In this sense, we also consider the definition of a typically non- or weakly informative prior for $\boldsymbol{x}$ pragmatically as part of the solution method.

For a clear understanding, the \textit{random forward problem} is stated as follows: Given $\boldsymbol{M}$, $\boldsymbol{A}$ and $\boldsymbol{B}$, how does a given input $\boldsymbol{x}$ map to outputs $\boldsymbol{y}$? 
We differentiate between three observation scenarios:
\begin{itemize}
\item[(i)] We assume to have independently observed output samples $\hat{\boldsymbol{y}}_l$ with individual observational uncertainties $\boldsymbol{B}_l$ as well as individual model parameters $\boldsymbol{A}_l$ ($l=1,\dots,L$) which leads to different outputs even for the same deterministic input $\boldsymbol{x}$, i.e. $\hat{\boldsymbol{y}}_l = \boldsymbol{M}(\boldsymbol{x};\boldsymbol{A}_l)+\boldsymbol{B}_l$. 
\item[(ii)] Contrary to scenario (i) we assume a single true model parameter set for all observations, but which is still a realization of an unknown parameter variable $\boldsymbol{A}$, i.e. $\hat{\boldsymbol{y}}_l = \boldsymbol{M}(\boldsymbol{x};\boldsymbol{A})+\boldsymbol{B}_l$. 
\item[(iii)] In this scenario we are observing not samples or realizations, but an output density function $f_{\hat{\boldsymbol{y}}}$ which contains also all observation uncertainties, i.e. $\hat{\boldsymbol{y}} = \boldsymbol{M}(\boldsymbol{x};\boldsymbol{A})$.
\end{itemize}
We assume all observation random variables as \textit{independent} with respect to $l$, such as $\boldsymbol{B}_l$ in scenarios (i) and (ii), and specifically in scenario (i) all $\boldsymbol{A}_l$ are \textit{i.i.d.} with the density $f_{\boldsymbol{A}}$.

In this work we are focusing on two sources of ambiguities in the inverse problem: First, since $\boldsymbol{M}$ is a general non-invertible function, a single $\boldsymbol{y}$ may be the result of many $\boldsymbol{x}$ just by the non-injectivity, which we call the \textit{structural ambiguity} of the inverse problem. Second, using mixture model random variables with separated components in $\boldsymbol{A}$ this additionally introduces a \textit{probabilistic ambiguity} in $\boldsymbol{y}$ under the condition that $\boldsymbol{M}$ propagates these mixture model components. This probabilistic ambiguity may contain a combinatorial side since all mixture model components of different parameters interact with each other for the calculation of $\boldsymbol{y}$. 

\textit{Remark:} This places $\boldsymbol{A}$ conceptually between the properties of $\boldsymbol{M}$, with ambiguities of the solution, and $\boldsymbol{B}$, which introduces uncertainty into the inverse problem. We consider the investigation of inverse problems with such random parameters $\boldsymbol{A}$ as a perspective that is not emphasized in the literature.

\subsection{General Solution with Random Equation Formalism}
\label{sec:GenSolForm}

The derivations are described by random equations and we investigate the likelihood of the difference random variable $\boldsymbol{M}(\boldsymbol{x};\boldsymbol{A})+\boldsymbol{B}-\boldsymbol{y}$ at position $\boldsymbol{0}$ similar to \cite{Hoegele 2026}. This is an alternative formulation to standard Bayesian inversion, e.g. \cite{Dashti 2017}, starting from an equation perspective. The derivations share the idea of Bayesian hierarchical models, e.g. see \cite{Sanz-Alonso 2025,Saibaba 2019,Gelman 2014}, using latent variable marginalization, but the main difference is that the latent variables are marginalized not as parameters of another density but of a general forward model, i.e. a hierarchical forward model. To compare it to a recent publication on hierarchical Bayesian inverse problems \cite{Sanz-Alonso 2025}, a difference is that in our inverse problem formulation we do not marginalize the prior with hyperpriors, but directly insert the latent variable nonlinearly inside the forward model as part of the likelihood function. We differentiate these likelihood derivations according to the three observation scenarios:\medskip

\begin{itemize}
\item[(i)] Since we have independent samples $\hat{\boldsymbol{y}}_l$ ($l=1,\dots,L$), we construct a likelihood function by (assuming stochastic independence of all $\boldsymbol{A}_l$ and $\boldsymbol{B}_l$, $l=1,\dots,L$)
\begin{align}
\mathcal{L}^{(i)}(\boldsymbol{0}|\boldsymbol{x}) = f_{\left\lbrace\boldsymbol{M}(\boldsymbol{x};\boldsymbol{A}_l)+\boldsymbol{B}_l-\hat{\boldsymbol{y}}_l\right\rbrace_{l=1}^{L}}(\boldsymbol{0}) = \prod\limits_{l=1}^L\;  f_{\boldsymbol{M}(\boldsymbol{x};\boldsymbol{A}_l)+\boldsymbol{B}_l-\hat{\boldsymbol{y}}_l}(\boldsymbol{0})\;.
\end{align}
Using the law of total probability for densities we further get (in the last step assuming that all entries of the observational random variable vector $\boldsymbol{B}_l$ are mutually independent)
\begin{align}
& = \prod\limits_{l=1}^L\;  \int\limits_{\mathbb{R}^K}  f_{\boldsymbol{M}(\boldsymbol{x};\boldsymbol{s})+\boldsymbol{B}_l-\hat{\boldsymbol{y}}_l}(\boldsymbol{0}) \cdot f_{\boldsymbol{A}}(\boldsymbol{s})\,\text{d}\boldsymbol{s}\\
& = \prod\limits_{l=1}^L\;\int\limits_{\mathbb{R}^K}   f_{\boldsymbol{B}_l}(\hat{\boldsymbol{y}}_l - \boldsymbol{M}(\boldsymbol{x};\boldsymbol{s})) \cdot f_{\boldsymbol{A}}(\boldsymbol{s})\,\text{d}\boldsymbol{s}\\
& = \prod\limits_{l=1}^L\;\int\limits_{\mathbb{R}^K}  \prod\limits_{r=1}^R\;   f_{B_{l,r}}(\hat{y}_{l,r} - M_r(\boldsymbol{x};\boldsymbol{s})) \cdot f_{\boldsymbol{A}}(\boldsymbol{s})\,\text{d}\boldsymbol{s}\;.\label{equ:likelihood}
\end{align}
This derivation in the random equation framework leads to a marginal likelihood representation with the latent random variables $\boldsymbol{A}_l$. 

\item[(ii)] Applying the law of total probability for densities first for a global latent variable $\boldsymbol{A}$ we get (again in the last step assuming that all entries of the observational random variable vector $\boldsymbol{B}_l$ are mutually independent)
\begin{align}
\mathcal{L}^{(ii)}(\boldsymbol{0}|\boldsymbol{x}) &= f_{\left\lbrace\boldsymbol{M}(\boldsymbol{x};\boldsymbol{A})+\boldsymbol{B}_l-\hat{\boldsymbol{y}}_l\right\rbrace_{l=1}^{L}}(\boldsymbol{0}) \\
& = \int\limits_{\mathbb{R}^K} f_{\left\lbrace\boldsymbol{M}(\boldsymbol{x};\boldsymbol{s})+\boldsymbol{B}_l-\hat{\boldsymbol{y}}_l\right\rbrace_{l=1}^{L}}(\boldsymbol{0}) \cdot f_{\boldsymbol{A}}(\boldsymbol{s})\,\text{d}\boldsymbol{s}\\
& = \int\limits_{\mathbb{R}^K}  \prod\limits_{l=1}^L\;    f_{\boldsymbol{M}(\boldsymbol{x};\boldsymbol{s})+\boldsymbol{B}_l-\hat{\boldsymbol{y}}_l}(\boldsymbol{0}) \cdot f_{\boldsymbol{A}}(\boldsymbol{s})\,\text{d}\boldsymbol{s}\\
& = \int\limits_{\mathbb{R}^K}  \prod\limits_{l=1}^L\;  f_{\boldsymbol{B}_l}(\hat{\boldsymbol{y}}_l - \boldsymbol{M}(\boldsymbol{x};\boldsymbol{s})) \cdot f_{\boldsymbol{A}}(\boldsymbol{s})\,\text{d}\boldsymbol{s}\\
& = \int\limits_{\mathbb{R}^K}  \prod\limits_{l=1}^L\;  \prod\limits_{r=1}^R\;   f_{B_{l,r}}(\hat{y}_{l,r} - M_r(\boldsymbol{x};\boldsymbol{s})) \cdot f_{\boldsymbol{A}}(\boldsymbol{s})\,\text{d}\boldsymbol{s}\;.\label{equ:likelihood_ii}
\end{align}
\end{itemize}
Equations (\ref{equ:likelihood}) and (\ref{equ:likelihood_ii}) handle the structural ambiguity through the non-invertibility of $\boldsymbol{M}$ and probabilistic ambiguity through the marginalizations over $\boldsymbol{A}_l$ or $\boldsymbol{A}$ respectively. It is not necessary that all $f_{\boldsymbol{B}_l}$ are identically distributed allowing for different observational error types. 

\begin{itemize}
\item[(iii)] If no observation samples $\left\lbrace\hat{\boldsymbol{y}}_l\right\rbrace_{l=1}^{L}$ are available but we only observe an output density function $ f_{\hat{\boldsymbol{y}}}$ directly (which is associated with an observation random variable $\hat{\boldsymbol{y}}$) the derivation needs to be adapted, e.g. see \cite{Hoegele 2024 imag} for a practical application in computer vision. We start with the system equation $\hat{\boldsymbol{y}} = \boldsymbol{M}(\boldsymbol{x};\boldsymbol{A})$ assuming that observational uncertainty is part of $f_{\hat{\boldsymbol{y}}}$ and that $\hat{\boldsymbol{y}}$ has mutually independent entries, i.e. that the densities of all entries are independently observed. Similarly, we derive the likelihood function by the difference random variable $\boldsymbol{M}(\boldsymbol{x};\boldsymbol{A})-\hat{\boldsymbol{y}} $ being $\boldsymbol{0}$:
\begin{align}
\mathcal{L}^{(iii)}(\boldsymbol{0}|\boldsymbol{x}) &= f_{\boldsymbol{M}(\boldsymbol{x};\boldsymbol{A}) - \hat{\boldsymbol{y}}}(\boldsymbol{0}) \\
&= \int\limits_{\mathbb{R}^K}  f_{\boldsymbol{M}(\boldsymbol{x};\boldsymbol{s}) -\hat{\boldsymbol{y}}}(\boldsymbol{0}) \cdot f_{\boldsymbol{A}}(\boldsymbol{s})\,\text{d}\boldsymbol{s} \\
&= \int\limits_{\mathbb{R}^K}  f_{\hat{\boldsymbol{y}}}(\boldsymbol{M}(\boldsymbol{x};\boldsymbol{s})) \cdot f_{\boldsymbol{A}}(\boldsymbol{s})\,\text{d}\boldsymbol{s}\\
&= \int\limits_{\mathbb{R}^K} \prod\limits_{r=1}^R\;  f_{\hat{y}_r}(M_r(\boldsymbol{x};\boldsymbol{s})) \cdot f_{\boldsymbol{A}}(\boldsymbol{s})\,\text{d}\boldsymbol{s}\;.\label{equ:likelihood_iii}
\end{align}
In this context, in the recent literature \textit{stochastic inverse problems} (SIP) are presented which also observe a full density $f_{\boldsymbol{y}}$. The main difference is that SIPs perform a direct \textit{push-forward} of an input density $f_{\boldsymbol{x}}$ in order to directly obtain $f_{\boldsymbol{y}}$ \cite{Marcy 2022, Qin 2024}. We want to clarify that the derived posterior in this scenario is not the same as the solution of the SIP: The posterior leads to the most probable inputs that fit to the observed output density (i.e. an update of knowledge about $\boldsymbol{x}$) while SIPs ask for the full density of inputs that produce the full density of outputs using the forward model. 
\end{itemize}

Due to standard Bayesian reasoning we get the according posterior densities $\pi^{(i)}_{\boldsymbol{x}|\left\lbrace\hat{\boldsymbol{y}}_l\right\rbrace_{l=1}^{L}}(\boldsymbol{x})$, $\pi^{(ii)}_{\boldsymbol{x}|\left\lbrace\hat{\boldsymbol{y}}_l\right\rbrace_{l=1}^{L}}(\boldsymbol{x})$ and $\pi^{(iii)}_{\boldsymbol{x}|\hat{\boldsymbol{y}}}(\boldsymbol{x})$ by multiplication $\mathcal{L}(\boldsymbol{0}|\boldsymbol{x})\cdot \pi_{\boldsymbol{x}}(\boldsymbol{x})$ with a typically weakly informative prior $\pi_{\boldsymbol{x}}$ and a subsequent normalization to probability mass $1$. 

Notably, these formulas for the posterior are quite general in the stated assumptions, including a) arbitrary dimensions and number of samples $R,n,K,L\in\mathbb{N}$, b) a general nonlinear, non-invertible and only piecewise continuous $\boldsymbol{M}$ containing (nonlinear) model uncertainties and c) no restrictions about the type of density functions of $\boldsymbol{A}$ and $\boldsymbol{B}$, especially including mixture models for $\boldsymbol{A}$.

\subsubsection*{Consistency Discussion}

First, approaches $(i)$ and $(ii)$ obviously coincide if only one sample is observed at the output, i.e. $L=1$. Furthermore, in this special case, $(iii)$ also coincides with these two approaches if the observed density of $\hat{y}_r$ is a shifted observation uncertainty random variable $\hat{y}_{r}^{\ast} + B_r$, where $\hat{y}_{r}^{\ast}\in\mathbb{R}$ is the single observation sample with a zero-centered symmetric observation uncertainty random variable $B_r$.

Second, the difference between approaches $(i)$ and $(ii)$ is that in $(i)$, the marginalization is performed for each sample individually, i.e. for a given $\boldsymbol{x}$ for each individual sample the best match between observation and forward prediction is taken into account considering all possible parameter values. This is different for approach $(ii)$ where the whole ensemble of observations is simultaneously matched to the forward predictions for a given $\boldsymbol{x}$ considering all possible parameter values.   

Third, for the specific case where the output is a deterministic formula of the input $\boldsymbol{N}(\boldsymbol{x})$ with only an additive parameter vector $\boldsymbol{A}$ (with independent entries), i.e. $\boldsymbol{M}(\boldsymbol{x};\boldsymbol{A}):=\boldsymbol{N}(\boldsymbol{x})+ \boldsymbol{A}$, in the approach $(i)$ the marginalization becomes a convolution, i.e. 
\begin{align}
\mathcal{L}^{(i)}(\boldsymbol{0}|\boldsymbol{x}) = \prod\limits_{l=1}^L\; \prod\limits_{r=1}^R\;   f_{B_{l,r}\ast A_r}(\hat{y}_{l,r} - N_r(\boldsymbol{x}))\;,
\end{align} 
which is the classical Bayesian inverse problem setup without marginalization and the effective observation uncertainty $\boldsymbol{B}_l+\boldsymbol{A}_l$ for each observation. Notably, this is not the case for approach $(ii)$.

Fourth, also for approach $(iii)$ multiple, say $L$, density observations can be performed. In this case, it is most practical if these densities are conditionally independent under the condition of a fixed parameter set $\boldsymbol{A}$, which is in line with the reasoning of approach $(ii)$ and leads to a product over $L$ under the integral. But it is also possible to consider the situation where for each density observation a different parameter set is assumed, which is in line with the reasoning of approach $(i)$ and leads to a product over $L$ outside the integral. In order to keep the focus, we exclude this question from of the rest of the paper and consider only single density observations. 

\subsection{Numerical Solution with Monte Carlo Integration}
\label{sec:NumSolForm}

The main numerical task is to approximate the derived likelihood functions by \textit{Monte Carlo integration} with its general convergence rate $O(\frac{1}{\sqrt{P}})$ for $P$ samples. Since these integrations are all marginalizations along $f_{\boldsymbol{A}}$ we are utilizing $P$ independently drawn samples $\boldsymbol{s}_p \sim f_{\boldsymbol{A}}$ in the following. We again differentiate between the three scenarios:

\begin{itemize}
\item[(i)] Applying this to Equation (\ref{equ:likelihood}) leads for a given $\boldsymbol{x}\in\mathbb{R}^n$ to 
\begin{align}
\mathcal{L}^{(i)}(\boldsymbol{0}|\boldsymbol{x}) &\approx  \prod\limits_{l=1}^L\; \left( \frac{1}{P} \sum\limits_{p=1}^P   \prod\limits_{r=1}^R\;   f_{B_{l,r}}(\hat{y}_{l,r} - M_r(\boldsymbol{x};\boldsymbol{s}_p))  \right)\;. \label{equ:numlikelihood2}
\end{align}
For numerical efficiency it is convenient that the $\boldsymbol{s}_p$ are drawn once for all samples $\hat{\boldsymbol{y}}_l$. 

\item[(ii)] For Equation (\ref{equ:likelihood_ii}) this leads for a given $\boldsymbol{x}\in\mathbb{R}^n$ to
\begin{align}
\mathcal{L}^{(ii)}(\boldsymbol{0}|\boldsymbol{x}) &\approx \frac{1}{P} \sum\limits_{p=1}^P   \prod\limits_{l=1}^L \prod\limits_{r=1}^R\;   f_{B_{l,r}}(\hat{y}_{l,r} - M_r(\boldsymbol{x};\boldsymbol{s}_p))  \;. \label{equ:numlikelihood2_ii}
\end{align}

\end{itemize}
The numerical effort for scenarios (i) and (ii) can be quantified by the number of evaluations of $\boldsymbol{M}$ and $f_{B_{l,r}}$. First, having $J$ grid points $\boldsymbol{x}_j$ we need $P\cdot J$ evaluations of all $M_r$. These can be calculated once for all samples $L$ leading to $P\cdot J\cdot R$ stored scalar values. Second, we need $P\cdot J\cdot L\cdot R$ evaluations of $f_{B_{l,r}}$ in Equation (\ref{equ:numlikelihood2}) which can dominate the total computational cost. If evaluating $\boldsymbol{M}$ is expensive, such as the numerical solution of differential equations, then attention should be directed towards smart sampling strategies for $\boldsymbol{s}_p$.

\begin{itemize}
\item[(iii)] Using Monte Carlo integration in Equation (\ref{equ:likelihood_iii}) we get for a given $\boldsymbol{x}\in\mathbb{R}^n$
\begin{align}
\mathcal{L}^{(iii)}(\boldsymbol{0}|\boldsymbol{x})&\approx \frac{1}{P} \sum\limits_{p=1}^P \prod\limits_{r=1}^R\; f_{\hat{y}_r}(M_r(\boldsymbol{x};\boldsymbol{s}_p)) \;.
\label{equ:numlikelihood1_SIP}
\end{align}
\end{itemize}
The numerical effort for scenario (iii) is focused on evaluations of $f_{\hat{\boldsymbol{y}}}$, which leads by using $J$ grid points $\boldsymbol{x}_j\in\mathbb{R}^n$ to $P\cdot J$ evaluations of this density function. The main effort here is that $f_{\hat{\boldsymbol{y}}}$ is typically not given analytically but is itself sampled and therefore interpolation routines need to be applied which can be numerically challenging.

After this approximation of the likelihood functions we need to apply a weakly informative prior $\pi_{\boldsymbol{x}}$ to get to the according posteriors, e.g. this could be defined in the computational region of interest $\Omega\subset\mathbb{R}^n$ with $\frac{1}{\text{vol}(\Omega)}$ for $\boldsymbol{x}\in\Omega$ and $0$ else. In the practical implementation, the region $\Omega$ is evaluated on a grid with points $\boldsymbol{x}_j$ ($j=1,\dots,J$) and the normalization of the posterior can be performed approximately calculating the Riemann sum and dividing the posterior by this value in order to scale the posterior to probability mass 1. Of course also more informative priors can be used if it is appropriate, for example, by defining $\Omega$ as a problem-specific feasibility region for solutions or other approaches.

With respect to the numerical effort of all scenarios, the curse of dimensionality can appear in two different ways: First, the computation on a grid $\boldsymbol{x}_j\in\mathbb{R}^n$ gets very costly for large $n$ since we need a large $J$ to cover $\Omega$. This could be reduced by Maximum A Posteriori (MAP) estimation or more advanced methods \cite{Sun 2022}. Second, the necessary number of samples $\boldsymbol{s}_p\in\mathbb{R}^K$ also might get very costly for large $K$ in order to adequately cover all regions of $\mathbb{R}^K$ where $f_{\boldsymbol{A}}$ has non-negligible density values.

We finally summarize the computation of the posteriors in two algorithms: For scenarios (i) and (ii) we present Algorithm 1 and for scenario (iii) Algorithm 2.
\begin{table}[htbp]
\hrulefill\\
\textbf{Algorithm 1 for scenarios (i) and (ii)} (in short: Algorithms 1$^{(i)}$ and 1$^{(ii)}$):
\begin{itemize}
\setlength{\itemsep}{0pt}
\item[1.] Identify the forward model $\hat{\boldsymbol{y}}_l = \boldsymbol{M}(\boldsymbol{x};\boldsymbol{A}_l) + \boldsymbol{B}_l$ for scenario (i) and $\hat{\boldsymbol{y}}_l = \boldsymbol{M}(\boldsymbol{x};\boldsymbol{A}) + \boldsymbol{B}_l$ for scenario (ii) with the system function $\boldsymbol{M}$ and the parameter random variables $\boldsymbol{A}_l$ or $\boldsymbol{A}$ and $\boldsymbol{B}_l$.
\item[2.] Collect the output samples $\left\lbrace\hat{\boldsymbol{y}}_l\right\rbrace_{l=1}^{L}$.
\item[3.] Draw $P$ samples $\boldsymbol{s}_p$ from $f_{\boldsymbol{A}}$.
\item[4.] Evaluate either Equation (\ref{equ:numlikelihood2}) for scenario (i) or Equation (\ref{equ:numlikelihood2_ii}) for scenario (ii) for each given input vector $\boldsymbol{x}$ utilizing the fact that $M_r(\boldsymbol{x};\boldsymbol{s}_p)$ are evaluated independently of $l$. Repeat this evaluation on a grid for $\boldsymbol{x}_j\in\mathbb{R}^n$ ($j=1,\dots,J$) that includes all possible posterior intensities. 
\item[5.] Multiply this likelihood map with a weakly informative prior and normalize it in order to approximate $\pi^{(i)}_{\boldsymbol{x}|\left\lbrace\hat{\boldsymbol{y}}_l\right\rbrace_{l=1}^{L}}(\boldsymbol{x})$ or $\pi^{(ii)}_{\boldsymbol{x}|\left\lbrace\hat{\boldsymbol{y}}_l\right\rbrace_{l=1}^{L}}(\boldsymbol{x})$ respectively.
\end{itemize}
\hrulefill
\end{table}

\begin{table}[htbp]
\hrulefill\\
\textbf{Algorithm 2 for scenario (iii)}:
\begin{itemize}
\setlength{\itemsep}{0pt}
\item[1.] Identify the forward model $\hat{\boldsymbol{y}} = \boldsymbol{M}(\boldsymbol{x};\boldsymbol{A})$ with the system function $\boldsymbol{M}$ and the parameter random variable $\boldsymbol{A}$.
\item[2.] Obtain the output density function $f_{\hat{\boldsymbol{y}}}$.
\item[3.] Draw $P$ samples $\boldsymbol{s}_p$ from $f_{\boldsymbol{A}}$.
\item[4.] Evaluate Equation (\ref{equ:numlikelihood1_SIP}) for each given input vector $\boldsymbol{x}$. Repeat this evaluation on a grid for $\boldsymbol{x}_j\in\mathbb{R}^n$ ($j=1,\dots,J$) that includes all possible posterior intensities. 
\item[5.] Multiply this likelihood map with a weakly informative prior and normalize it in order to approximate $\pi^{(iii)}_{\boldsymbol{x}|\hat{\boldsymbol{y}}}(\boldsymbol{x})$.
\end{itemize}
\hrulefill
\end{table}

\section{Simulation Results}

\subsection{Analytical Quadratic Forward Models for Demonstration and Understanding}

Four illustrative models that exhibit structural and probabilistic ambiguities are given by (with $R=1$, $n,K\in\{1,2\}$)
\begin{align}
I) &\quad y = M(x;A) + B = A_1\cdot x^2 + B\\
II) &\quad y = M(x;\boldsymbol{A}) + B = A_1\cdot (x-A_2)^2 + B\\
III) &\quad y = M(\boldsymbol{x};A) + B = A_1\cdot (x_1^2 + x_2^2) + B\\
IV) &\quad y = M(\boldsymbol{x};\boldsymbol{A}) + B = A_1\cdot x_1^2 + A_2\cdot x_2^2 + B
\end{align}
which is essentially a random extension of a toy example \cite{Baattrupp 2026}. The random variable $\boldsymbol{A}$ contains mono- and multi-modal probability densities and $B\sim\mathcal{N}(0,0.1^2)$ is a Gaussian. The output samples $\hat{y}_l$ ($l=1,\dots,L$) for Algorithm 1 are drawn from the three random variables $A_1,A_2$ and $B$ individually for each $l$ simulating the \textit{random forward model}. For applying Algorithm 2, we generate a full output density $f_{y}$. For the calculation of all posteriors $P=10000$ draws of $f_{\boldsymbol{A}}$ applying latin hypercube sampling are utilized. 

\begin{figure}[htbp]
\centering
\includegraphics[width=16cm]{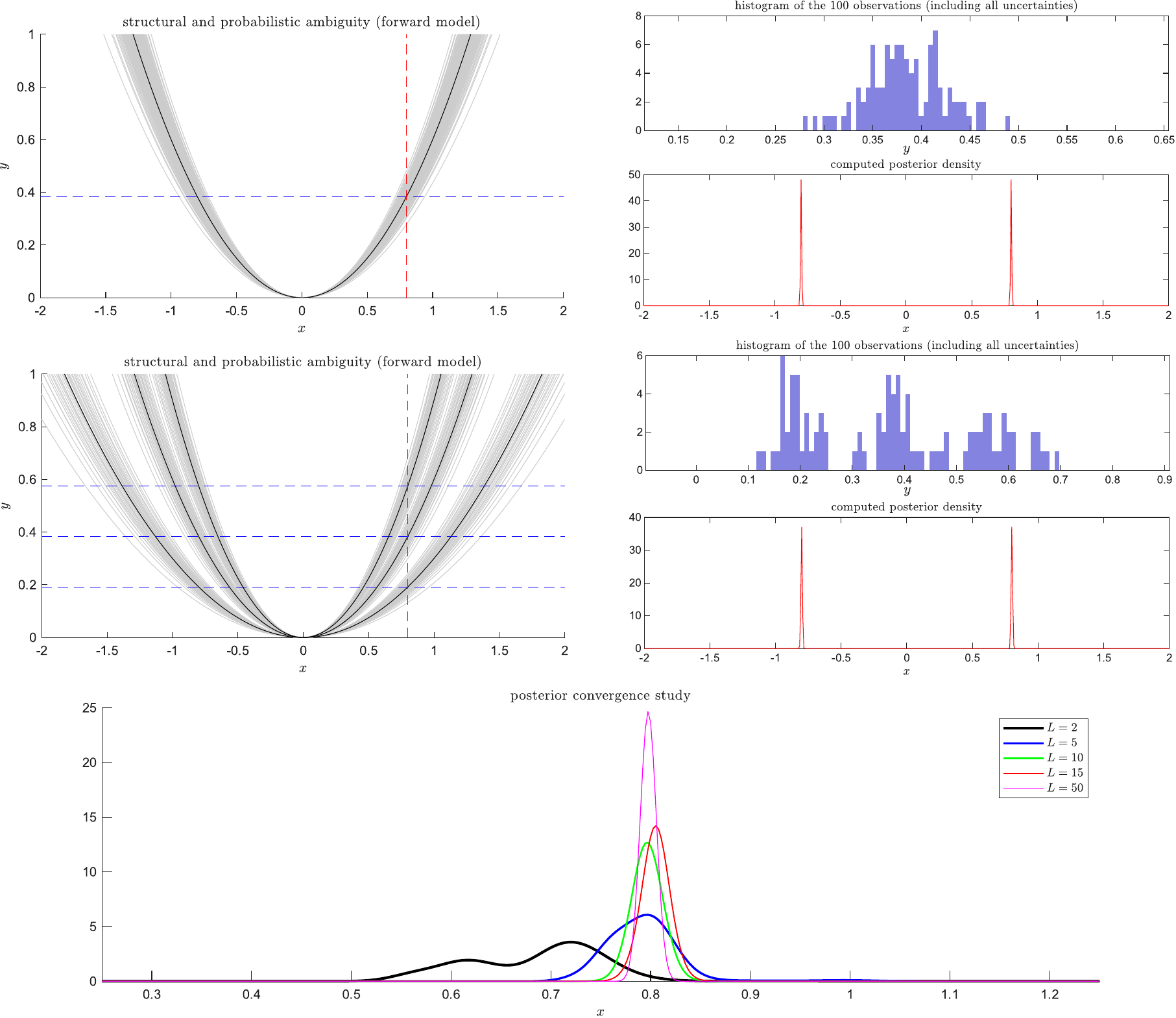}
\caption{Simulation results for the 1D quadratic random inverse problems $I$ using Algorithm 1$^{(i)}$. First row: with a mono-modal parameter random variable, second row: with a tri-modal parameter random variable. In the third row: convergence study utilizing the same setup as in the second row zoomed in around the right posterior peak with increasing number of observations $L$. }  
\label{fig:Res:Fig1}
\end{figure}

In Figure \ref{fig:Res:Fig1} simulations and posterior reconstructions for the 1D model $I$ are presented utilizing Algorithm 1$^{(i)}$, i.e. scenario $(i)$. The forward model can be understood by the plots on the left, where we show three cases of the quadratic function (first row: model $I$ with mono-modal $f_{A_1}$ with $\mathcal{N}(0.6,0.06^2)$, second row: model $I$ with tri-modal $f_{A_1}$ with equally weighted separated components $\mathcal{N}(0.3,0.04^2)$, $\mathcal{N}(0.6,0.06^2)$ and $\mathcal{N}(0.9,0.08^2)$, third row: same random variable setup as in the second row). All randomly drawn quadratic functions for the generation of the output are plotted as gray curves. The deterministic input value $x=0.8$ is shown as vertical red dashed line and the output values of the mixture model modes are presented as horizontal blue dashed lines. In each row on the top right $L=100$ output samples are shown as blue colored histogram (including the random observational error $B_l$).  On the bottom right the calculated posterior based on these samples are presented as red curves. The third row shows a posterior convergence study for sample sizes $L\in\{2,5,10,15,50\}$ zoomed in at the right posterior peak (the left peak is symmetric to the right). 
These results show: In the first row only structural ambiguity and a slight parameter blur is present with two probable solutions in the posterior at $x=0.8$ (true input) and the second solution $x=-0.8$. In the second row due to the probabilistic ambiguity the output samples show high complexity but the probabilistic ambiguity can be resolved in the posterior leading to the same posterior as in the first row. The convergence study in the third row shows increasing precision and accuracy around the true input value $x=0.8$ with increasing observation sample numbers.

\begin{figure}[htbp]
\centering
\includegraphics[width=16cm]{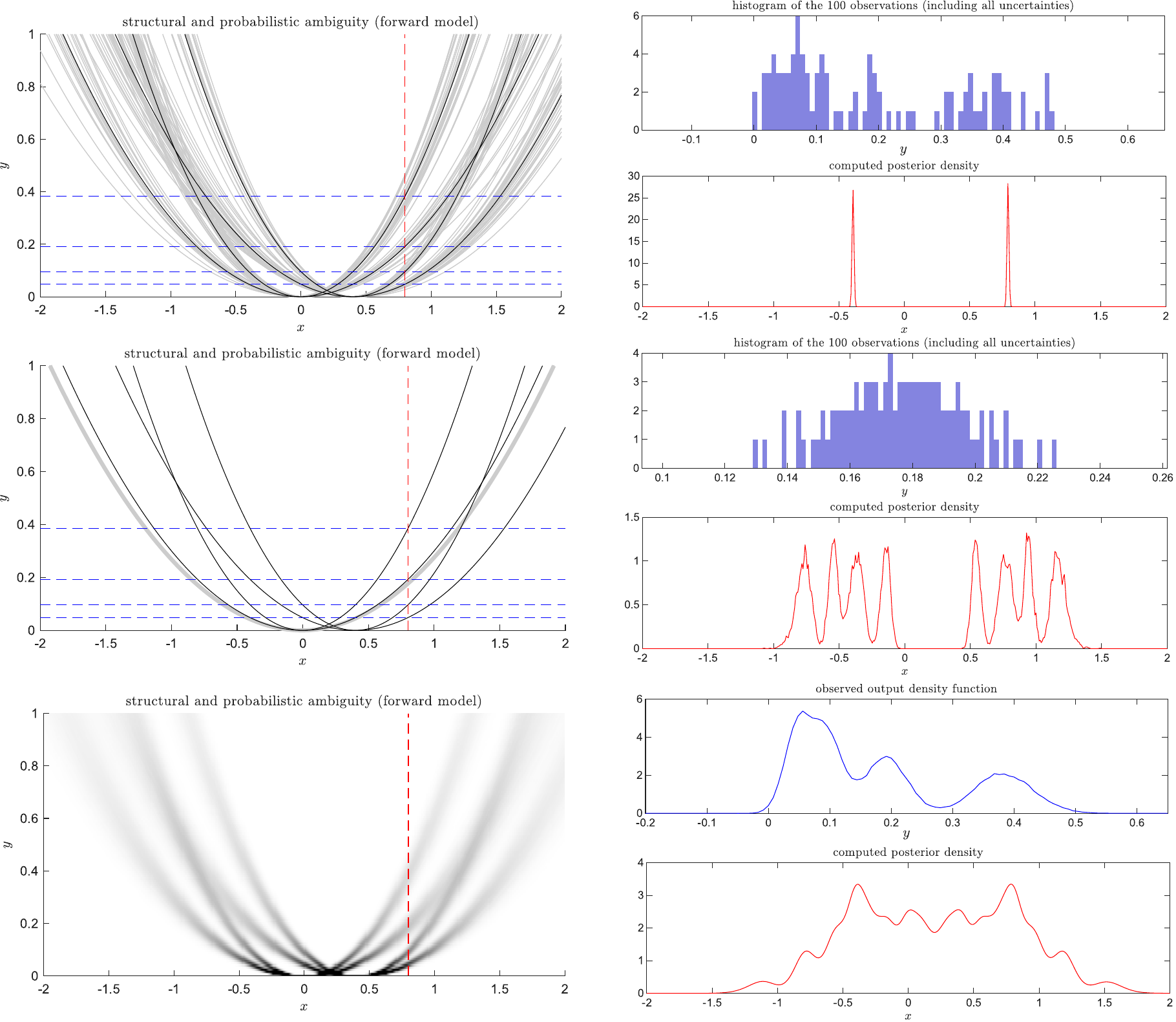}
\caption{Simulation results for the 1D quadratic random inverse problems $II$ with two bi-modal parameter random variables (leading to four combinations of quadratic functions at the parameter modes). First row: applying Algorithm 1$^{(i)}$, second row: applying Algorithm 1$^{(ii)}$ and third row: applying Algorithm 2.}  
\label{fig:Res:Fig2}
\end{figure}

In Figure \ref{fig:Res:Fig2} simulations and posterior reconstructions for the 1D model $II$ are presented comparing Algorithm 1$^{(i)}$, Algorithm 1$^{(ii)}$ and Algorithm 2, i.e. all three scenarios. The forward model uses in all scenarios bi-modal $f_{A_1}$ with equally weighted separated components $\mathcal{N}(0.3,0.04^2)$, $\mathcal{N}(0.6,0.06^2)$ and $f_{A_2}$ with $\mathcal{N}(0,0.02^2)$, $\mathcal{N}(0.4,0.02^2)$. The first row utilizing Algorithm 1$^{(i)}$ and second row utilizing Algorithm 1$^{(ii)}$ presentations are analogous to Figure \ref{fig:Res:Fig1}, except in the third row utilizing  Algorithm 2 a 2D histogram in the left is shown and on the top right the observed density function. On the bottom right in the first to third row the calculated posterior based on these samples or on the observed density are presented as red curves. 
The results show: In all plots additionally to scaling also a shift of the quadratic function is present. In the first row applying Algorithm 1$^{(i)}$ again the probabilistic ambiguity is fully resolved leading to two solutions of structural ambiguity but it changes the symmetry (shifting it away from $0$ compared to Figure \ref{fig:Res:Fig1}) - showing an interplay between both sources of ambiguity. In the second row applying Algorithm 1$^{(ii)}$ it is presented that the approach does not resolve structural and probabilistic ambiguities and their mixture (with eight distinct peaks consistent with the observed values $\hat{\boldsymbol{y}}_l$) is present in the posterior. This means, in this scenario the probabilistic ambiguity can backpropagate to the solution in input space. This can be understood directly since the marginalization in scenario (ii) is performed with only a single realization of $\boldsymbol{A}$ in the observations which makes different inputs $\boldsymbol{x}$ plausible depending with which mode combination of $\boldsymbol{A}$ in the observation this input is associated with. In the left plot this is directly observable: the plurality of mode combinations of quadratic functions (the black parabolas) leads to several plausible $x$-values leading to the observed $y$-value coming from the gray parabola. In the third row it can be observed that Algorithm 2 recovers the same posterior peaks as Algorithm 1$^{(i)}$ but they are much less distinct. This can be understood directly since observing an output density can be interpreted as a single sample observation with a characteristic but also quite broad uncertainty. This means many $x$ values are plausible leading to $y$ values with nonzero $f_y$ values.

\begin{figure}[htbp]
\centering
\includegraphics[width=14cm]{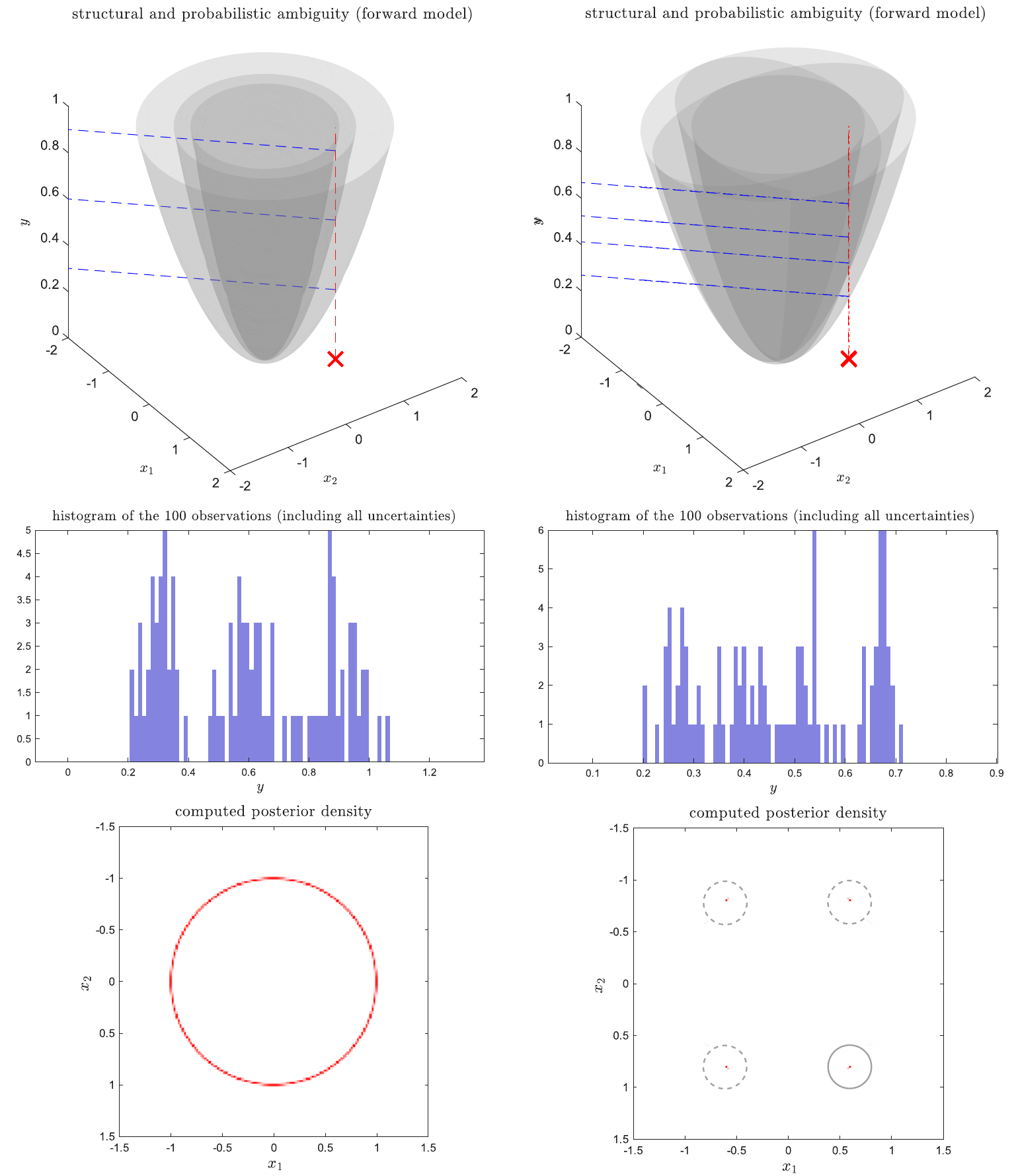}
\caption{Simulation results for the 2D quadratic random inverse problems $III$ and $IV$ applying Algorithm 1$^{(i)}$. First column: model $III$ with a tri-modal parameter random variable, second column: model $IV$ with two bi-modal parameter random variables (leading to four combinations of paraboloids at the parameter modes).}
\label{fig:Res:Fig3}
\end{figure}

In Figure \ref{fig:Res:Fig3} simulations and posterior reconstructions for the 2D models $III$ and $IV$ are presented applying Algorithm 1$^{(i)}$. The plot structure is similar as for Figure \ref{fig:Res:Fig1} in each column: on the top the paraboloids of the mixture model modes are shown, in the center the observed samples and on the bottom the computed 2D posterior intensity maps. The used densities are: in the first column, model $III$ with tri-modal $f_{A_1}$ with equally weighted separated components $\mathcal{N}(0.3,0.04^2)$, $\mathcal{N}(0.6,0.06^2)$ and $\mathcal{N}(0.9,0.08^2)$ and in the second column, model $IV$ with bi-modal $f_{A_1}$ with equally weighted separated components $\mathcal{N}(0.3,0.06^2)$, $\mathcal{N}(0.7,0.04^2)$ and $f_{A_2}$ with $\mathcal{N}(0.25,0.02^2)$, $\mathcal{N}(0.65,0.02^2)$. The deterministic input value is $x=(0.6,0.8)$.
The results show: In the first column, the full circular posterior (which fits to the observed samples) is computed due to radial symmetry as structural ambiguity of the model which contains the true input value. This illustrates that the method can represent 2D posteriors with infinitely many solutions. In the second column, a mixture of four differently scaled paraboloids is utilized which breaks radial symmetry but still contains four possible points in the posterior representing the residual structural ambiguity. Again, all probabilistic ambiguities are resolved in Algorithm 1$^{(i)}$.

\subsection{Advanced Numerical Forward Model I: Epidemiological Inverse Parameter Estimation using the SIR-Model}

\begin{figure}[htbp]
\centering
\includegraphics[width=16cm]{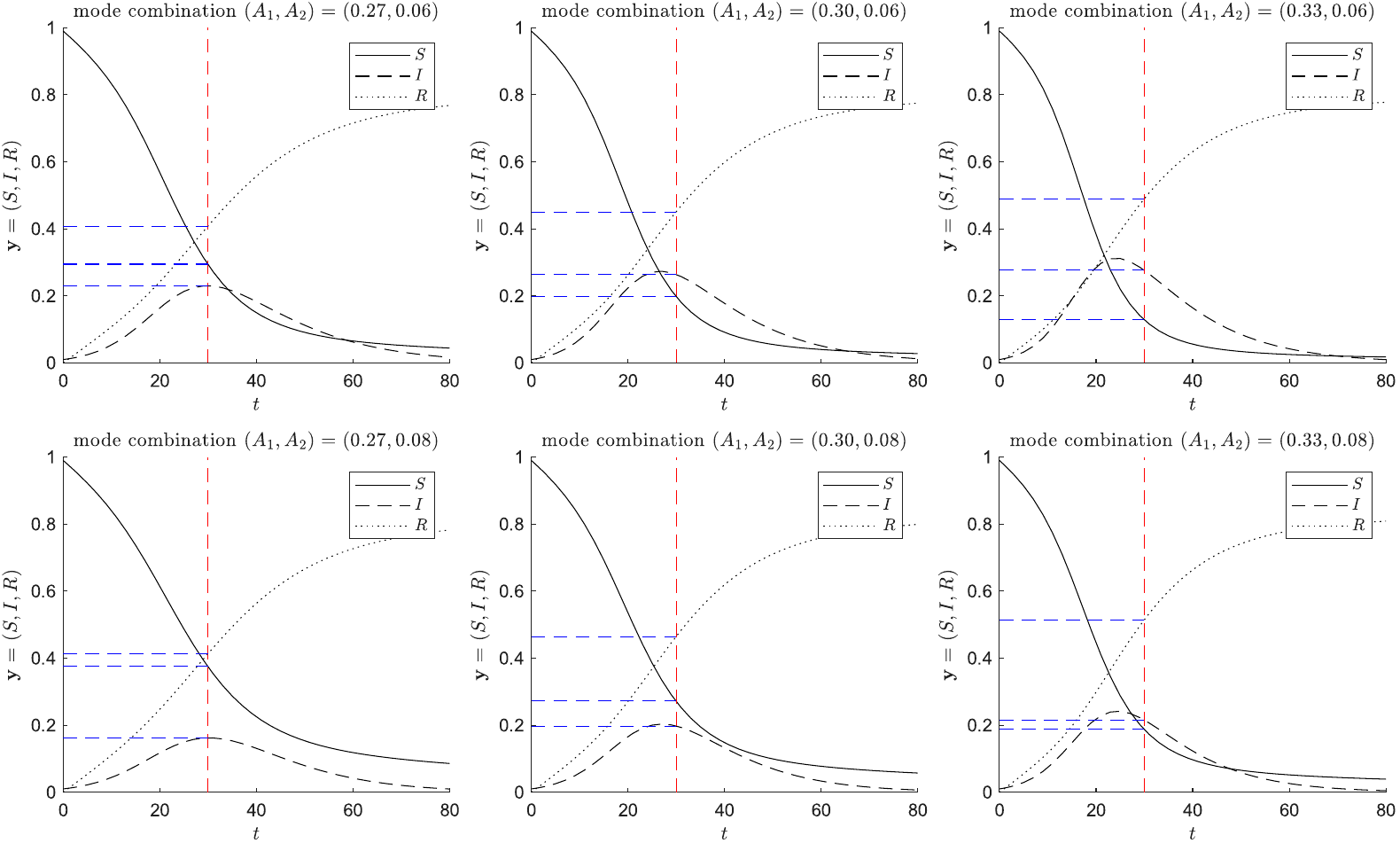}
\caption{Simulation results for the SIR forward model for the six parameter mode combinations and the measurements without noise (blue dashed lines).}
\label{fig:Res:Fig4}
\end{figure}

In this numerical example we are focusing on the influence of the probabilistic ambiguity for a variation of the epidemiological SIR model as presented recently \cite{Chada 2026} for Bayesian inverse problems with respect to unbiased estimators. The dynamical system is described by the following ODE system
\begin{align}
\dot{S}(t) &= -A_1\,S(t)\,I(t) - x_1\,S(t)\\
\dot{I}(t) &= A_1\,S(t)\,I(t) - (A_2 + x_1 + x_2)\,I(t)\\
\dot{R}(t) &= A_2\,I(t) + x_1\,S(t)\;,
\end{align}
with $S$ the proportional factor of susceptible parts of the population to the disease, $I$ the proportion of infected individuals and $R$ the recovered and immune proportion of individuals. The multi-modal parameters with uncertainty are given by $\boldsymbol{A}$ and the parameters that need to be found by the inverse problem are $\boldsymbol{x}$. The initial conditions of the system are chosen to $(S(0),I(0),R(0)) = (0.99,0.01,0)$. The forward model function is described by observations of these three populations at a specific time $t=30$ where the system shows a high dynamic
\begin{align}
\boldsymbol{M}(\boldsymbol{x};\boldsymbol{A}) = \left( \begin{array}{c} S(30) \\ I(30) \\ R(30) \end{array} \right)\;.
\end{align}
This represents a higher-dimensional use case of this methodology with problem parameters $R=3,K=2,n=2$. For the numerical investigations we use the following mixture models $A_1$ a tri-modal $f_{A_1}$ with equally weighted separated components $\mathcal{N}(0.27,0.01^2)$, $\mathcal{N}(0.3,0.01^2)$ and $\mathcal{N}(0.33,0.01^2)$, $A_2$ a bi-modal $f_{A_2}$ with equally weighted separated components $\mathcal{N}(0.06,0.01^2)$ and $\mathcal{N}(0.08,0.01^2)$, for the observation noise $f_{B_1} = f_{B_2} = f_{B_3} = \mathcal{N}(0,0.005^2)$ and the true parameter values $x_1 = 0.01,x_2=0.01$. In total, this leads to six mode combinations of the parameters $\boldsymbol{A}$ which are presented in Figure \ref{fig:Res:Fig4} with the measurements without noise for these modes shown as blue dashed lines.

\begin{figure}[htbp]
\centering
\includegraphics[width=16cm]{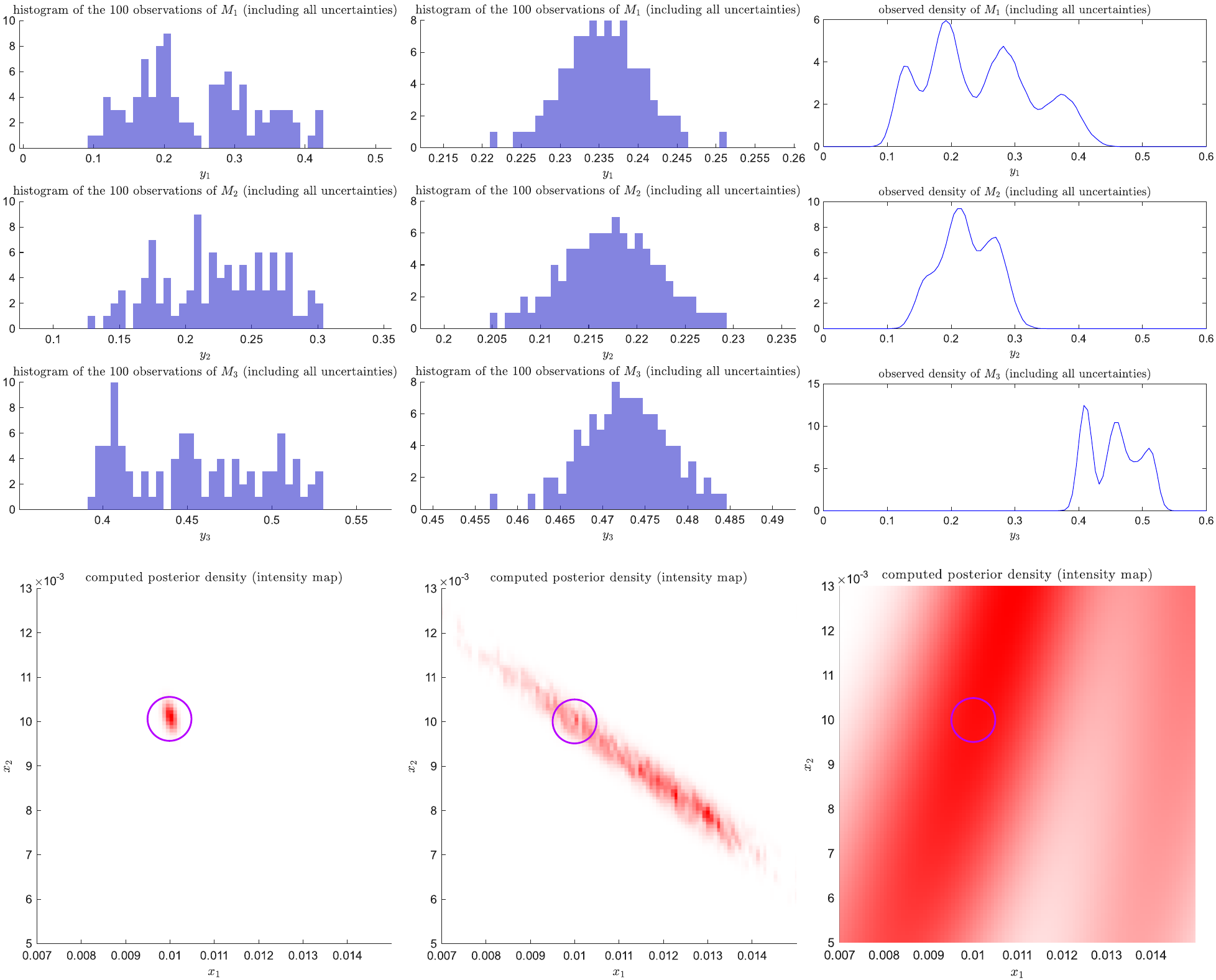}
\caption{Posterior computation of the inverse parameter estimation for the SIR model. In the three columns the results for scenario $(i)$ to $(iii)$ are presented. In the top three rows the three observation dimensions for each scenario are presented in blue color. In the bottom row the computed posteriors are presented in red color as intensity maps and the true input position as violet circle.}
\label{fig:Res:Fig5}
\end{figure}

In Figure \ref{fig:Res:Fig5} three columns of the results for observation scenarios $(i)$, $(ii)$ and $(iii)$ are presented. In the top three rows the observations of $\boldsymbol{y}_l$ ($l=1,\dots,L=100$) for the first two columns and the observed densities of $\boldsymbol{y}$ are presented in the third column.  For the simulations of observation scenario $(i)$ we use $P=1000$ samples of $\boldsymbol{A}$ and $J=101\times 101$ data points for a 2D grid for $\boldsymbol{x}$, leading to $10,201,000$ simulations of the ODE system and for scenario $(ii)$ for $J=4410$ data points with $P=90000$ were used, leading to $396,900,000$ simulations of the ODE system which was solved using a Runge Kutta 2-3 scheme. The bottom row shows the derived posteriors as a 2D intensity map. In this example no structural ambiguity is present as can be observed in the left column for scenario $(i)$ and we focus on the probabilistic ambiguity. 

The posterior for scenario $(i)$ shows a distinct peak a the true position of the parameters and this is in agreement to the previous observation that the probabilistic ambiguities can be fully resolved in this scenario. For scenario $(ii)$ the computation of the posterior took a considerable computational effort and the resulting intensity map shows a characteristic shape which is on a distinct line but on that line there is a broad intensity spread with the highest intensities not at the true $(x_1,x_2)$ position indicating a bias. In this example the probabilistic ambiguities are not getting fully visible compared to the example before but the overall assessment of a backpropagation of model uncertainty into the posterior is in agreement. The posterior for scenario $(iii)$ shows high intensities at the correct position but the peak shape is very broad being not as distinct as compared to scenario $(i)$.

In total, this more advanced example leads to the same conclusions as the simpler quadratic example.

\subsection{Advanced Numerical Forward Model II: Inverse Source Localization for the Heat PDE with Multi-Modal Conductivity}

\begin{figure}[htbp]
\centering
\includegraphics[width=7cm]{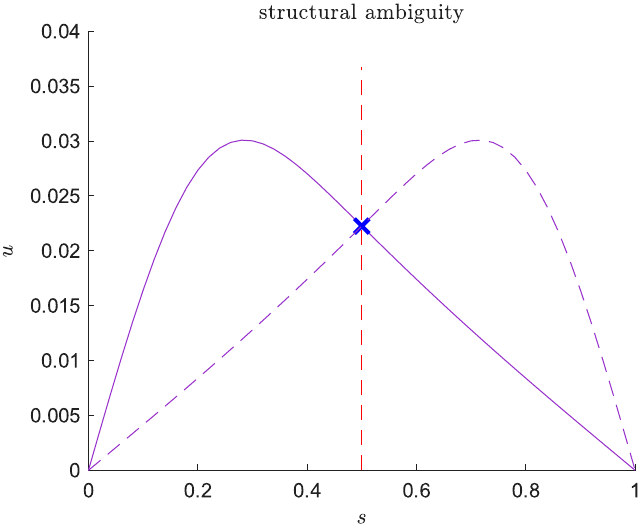}
\caption{Demonstration of the structural ambiguity for the 1D heat PDE with central measurement. Shown is an example heat distribution with an initial source position at $x=0.2$ (solid line) and it is presented that the same measurements would have been obtained for a source at position $x=0.8$ (dashed line) due to symmetry of the problem definition.}
\label{fig:Res:Fig6}
\end{figure}

In this more advanced inverse problem we are focusing on the 1D heat transfer partial differential equation which describes a diffusion process, e.g. in a wire, and which will be heated by a thermal source. The goal is to solve the inverse heat source problem, i.e. determining the source position by utilizing measured heat values. This is a challenging standard problem in applied mathematics \cite{Saibaba 2019,El Badia 2002,Yang 2011} and the general PDE is given by
\begin{align}
u_t - A\cdot u_{ss} = q(x,s,t)\;,
\end{align}
with $u(s,t)$ the heat function depending on the space coordinate $s$ and time coordinate $t$, $A\in\mathbb{R}$ the material dependent conductivity coefficient and $q(x,s,t)$ the thermal source term which depends on the source position $x$. In this demonstration we set $s\in[0,1]$ and $t\in[0,2]$. In particular, we assume a source function of type
\begin{align}
q(x,s,t) := e^{-\frac{(s-x)^2}{2\,\sigma_q^2}}\,\varepsilon(t-t_0)\;,
\end{align}
where $x$ describes the position of the highest intensity of a Gaussian heat source and $\varepsilon(t-t_0)$ the shifted Heaviside function which essentially means that the thermal source is turned on in $[0,t_0]$ and then turned off. We will use $t_0=1.17$ as the standard heating time, $\sigma_q=0.02$ as the standard deviation (=width) of the heat source, the initial condition $u(s,0)=0$ and the boundary conditions $u(0,t)=u(1,t)=0$. This means, until $t_0$ the wire is heated from constant zero temperature with a Gaussian source at $x$ and after that time it is cooling down again due to the boundary conditions. For given $A$ and $x$ the concrete heat function $u(s,t)$ can computed numerically utilizing FEM in Matlab with the \textit{pdepe}-solver. The forward model value is a point measurement of this heat function. This point measurement is chosen to introduce structural ambiguities for the inversion, e.g. a measurement at the center of the spatial coordinate $s=0.5$ directly after the heating phase $t=1.2$, i.e.
\begin{align}
M(x;A) := u(0.5,1.2)\;. 
\end{align}
This structural ambiguity can be directly understood in Figure \ref{fig:Res:Fig6} where for an asymmetric source position $x$ the measurement (blue cross) will be the same also for source position $1-x$ (comparing the solid line with the dashed line). In total, this leads to the problem parameters $R=K=1$, $n\in\{1,2\}$ where the case $n=2$ is presented later. 

\begin{figure}[htbp]
\centering
\includegraphics[width=16cm]{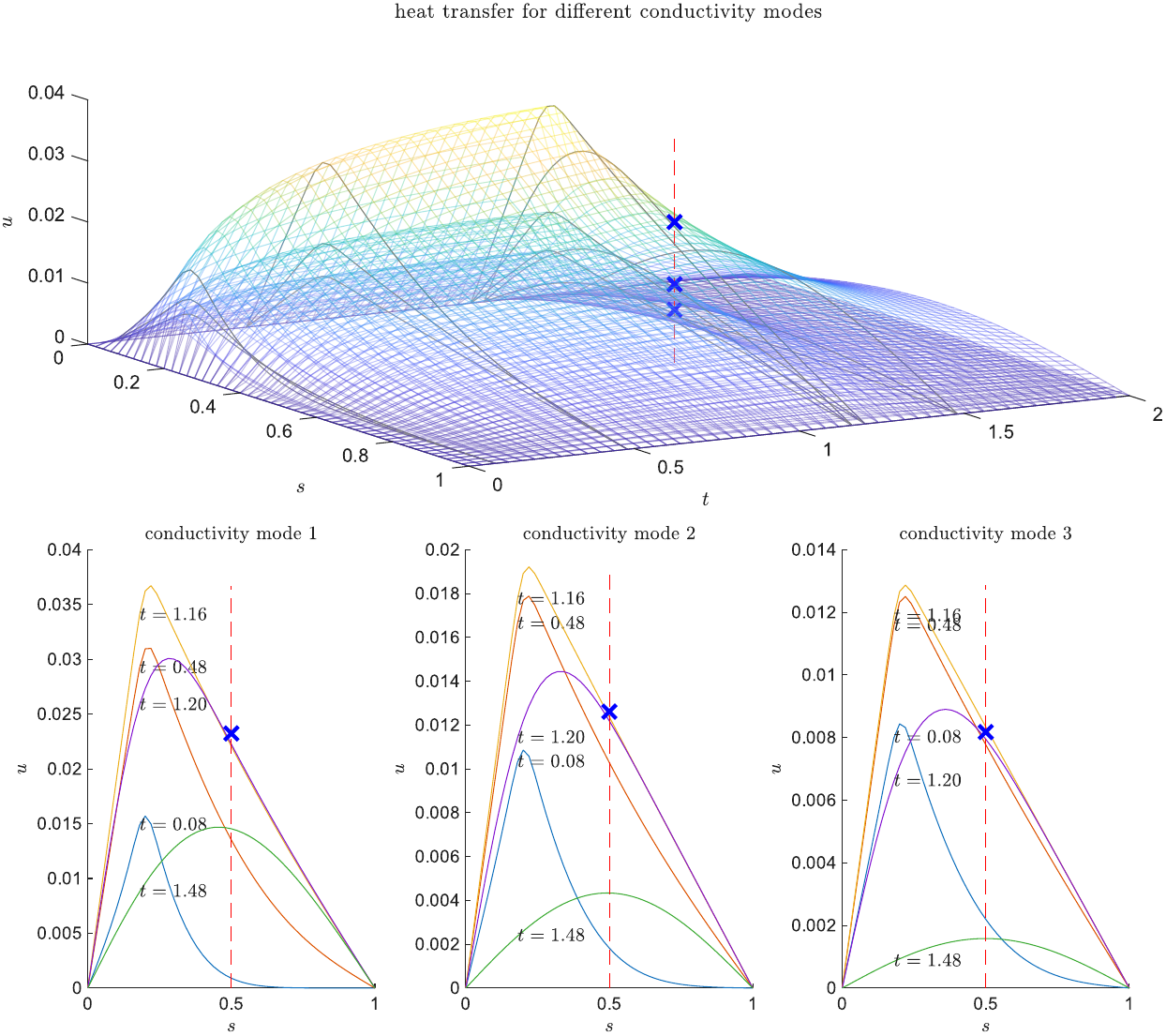}
\caption{Simulation results for the 1D heat transfer PDE forward model with an asymmetric heat source. Top: the graph of the 2D heat function depending on space $s$ and time $t$ coordinates for the three conductivity modes. Additionally the measurement conducted at $s=0.5$ and $t=1.2$ is presented as red dashed lines and the nominal values at the three conductivity modes are presented as blue crosses, representing the probabilistic ambiguity. Bottom: row from left to right: slices of the heat distribution above at discrete time points for each conductivity mode separated.}
\label{fig:Res:Fig7}
\end{figure}

In the simulations we will use for the conductivity $A$ a tri-modal $f_{A}$ with equally weighted separated components $\mathcal{N}(0.2,0.02^2)$, $\mathcal{N}(0.4,0.04^2)$ and $\mathcal{N}(0.6,0.06^2)$, for the observation noise $f_{B} = \mathcal{N}(0,0.002^2)$ and the true source position $x=0.2$. The tri-modal conductivity essentially means that it is unclear which observation belongs to which conductivity regime and despite of this probabilistic ambiguity we want to determine the source position $x$.  In Figure \ref{fig:Res:Fig7} the simulation of the forward model of the heat transfer and the undisturbed measurement values are presented for the three conductivity modes. 

\begin{figure}[htbp]
\centering
\includegraphics[width=16cm]{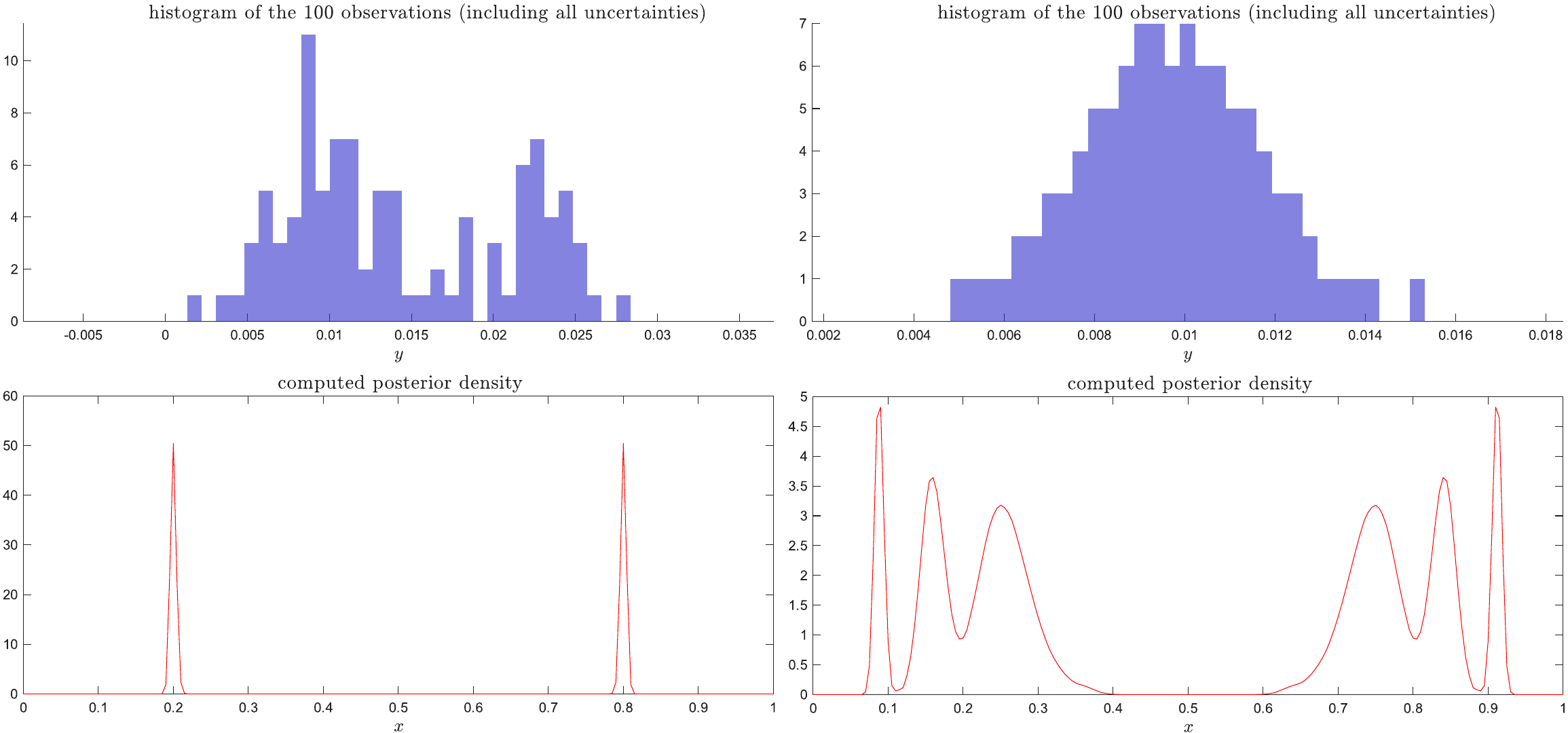}
\caption{Solution of the heat transfer PDE inverse model with the two observation scenarios $(i)$ (left column) and $(ii)$ (right column). Top: the histogram of the observations with system parameters drawn from the multi-modal parameter distribution for each sample (left) or with one true parameter set (right). Bottom: The resulting posterior distributions with only residual structural ambiguities (left) and an unresolved mixture of structural and probabilistic ambiguities (right).}
\label{fig:Res:Fig8}
\end{figure}

In Figure \ref{fig:Res:Fig8} the results for observation scenario $(i)$ (left) and scenario $(ii)$ (right) are presented. The results show that the probabilistic ambiguity of the measurement could be resolved for Algorithm 1$^{(i)}$ with the posterior showing distinct peaks at $x=0.2$ and $x=0.8$.  On the right, Algorithm 1$^{(ii)}$ also shows distinct peaks but they are practically not resolved, i.e. the main peak position are slightly shifted compared to the true value and there are six peaks (coming from two peaks due to structural ambiguity times three peaks due to the tri-modal probabilistic ambiguity). We regard this as degraded results compared to Algorithm 1$^{(i)}$ which is in accordance to the observations of the previous simulation results. $P=1000$ samples of $A$ were used for the computation of the posterior at $J=201$ data points, leading to $201,000$ simulations of the 1D heat equation in total.

\begin{figure}[htbp]
\centering
\includegraphics[width=16cm]{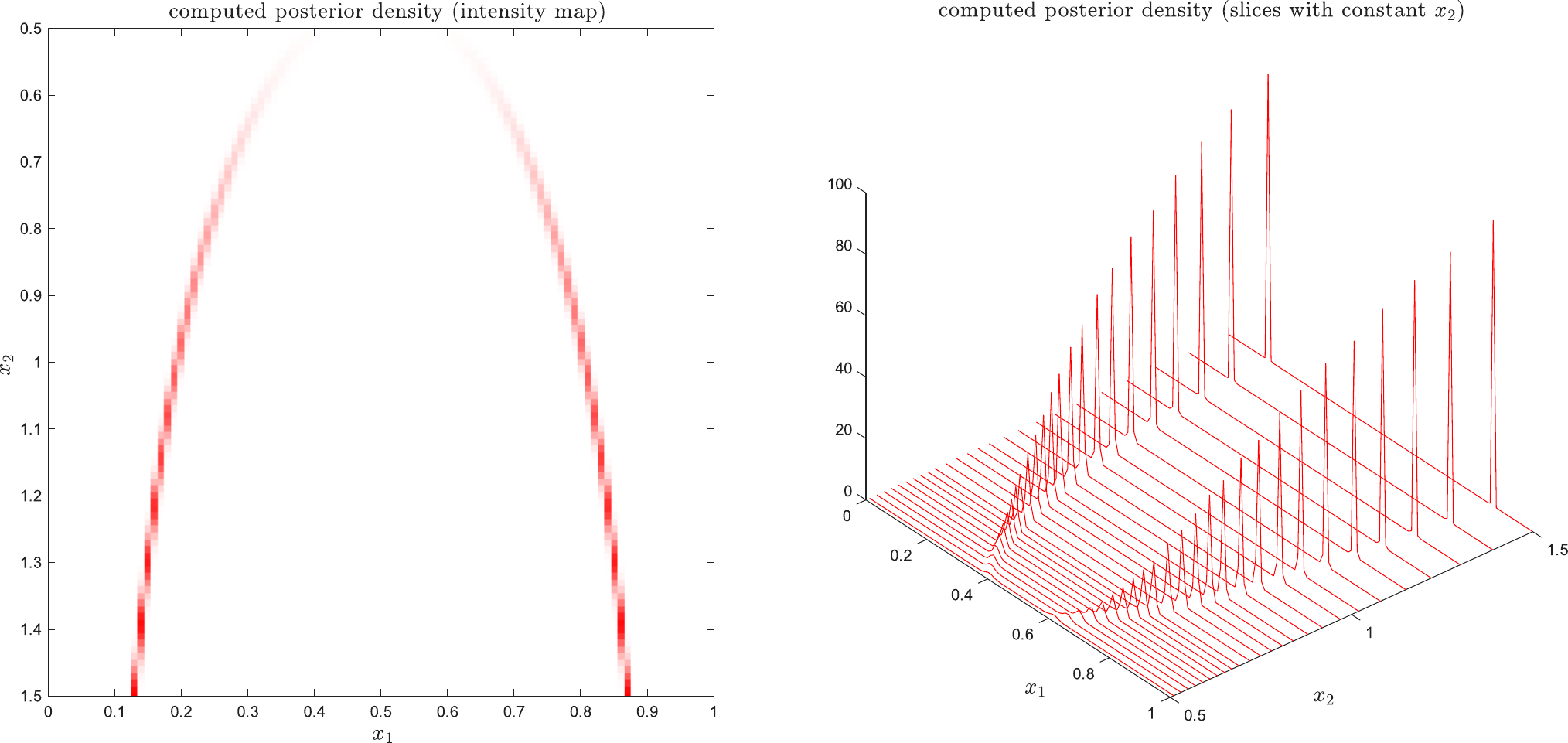}
\caption{Solution of the heat transfer PDE inverse model for the observation scenario $(i)$ and for the source with two parameters: $x_1$ the position and $x_2$ the intensity of the heat source. The resulting 2D posterior distribution with only residual structural ambiguities is presented as intensity map (left) and as slice with constant $x_2$ values (right).}
\label{fig:Res:Fig9}
\end{figure}

We further consider a modified heat source term
\begin{align}
q(\boldsymbol{x},s,t) := x_2\cdot e^{-\frac{(s-x_1)^2}{2\,\sigma_q^2}}\,\varepsilon(t-t_0)\;,
\end{align}
where $x_1$ is still describing the position of the heat source but additionally $x_2$ describes the intensity of the heat source. In the following all other problem parameters are kept as before. Obviously there is a much richer structural ambiguity present since the same measurement value can be observed for higher intensities (i.e. $x_2>1$) at heat source positions that are further away from $s=0.5$ than the true value (i.e. $|x_1-0.5|>0.3$) as for lower intensities (i.e. $x_2<1$) closer to $s=0.5$ (i.e. $|x_1-0.5|<0.3$). A nonlinear structural inverse source problem characteristic can be expected showing this trade off and the corresponding 2D posterior for the observation scenario $(i)$ is presented in Figure \ref{fig:Res:Fig9} quantifying this characteristic (based on very similar output observations as in Figure \ref{fig:Res:Fig8} on the top left). For this simulation $J=101\times 101$ data points for $\boldsymbol{x}$ were used, leading to the considerable amount of $10,201,000$ simulations of the 1D heat equation. Again the probabilistic ambiguities are resolved showing only residual structural ambiguities in the 2D posterior and the 1D posterior of Figure \ref{fig:Res:Fig8} bottom left is a slice at $x_2=1$.

This example indicates that this methodology can be applied to forward models which can only be calculated numerically such as inverse source localization problems based on PDEs.

\section{Discussion}
\label{sec:discussion}

The goal of this work is to broaden the perspective on uncertainty-driven inverse problems by investigating \textit{random inverse problems}. As a main contribution we subdivide ambiguities of the inverse problem as structural, due to non-injectivity of the forward problem, and probabilistic, due to multi-modal mixture model parameters which are nonlinearly in the forward model. The latter ambiguity can be regarded as considering a stochastic family of forward models with concentrated modes. A central observation is that these two types of ambiguities interact with each other, depending on observation scenarios at the output. This perspective extends the classical Bayesian inversion presentation since it treats deterministic forward problems with classical additive noise without a concept of probabilistic ambiguities.\medskip

In the simulation results for the inversion, e.g. for 1D and 2D quadratic functions which contain structural ambiguities by design, we distinguish between three observation scenarios: (i) For each observation not only random noise is newly sampled but also the parameter random variable which is a fundamental uncertainty in the forward problem. It is demonstrated that this scenario makes it possible that the probabilistic ambiguities can be resolved leaving only structural ambiguities in the posterior. (ii) In this scenario there is one true but unknown parameter set and only observational noise is sampled for each observation. This leads to a backpropagation of the multi-modal uncertainty about the parameters in the posterior, showing a practically irresolvable combination of structural and probabilistic ambiguities. (iii) Contrary to the previous scenarios, not samples but a full output density is observed. In this case probabilistic ambiguities can also be resolved but with a much lower credibility level compared to (i). Additionally, an epidemiological SIR ODE model inverse parameter estimation problem and a heat PDE model the inverse source localization problem with structural and probabilistic ambiguities are solved in this framework showing the same characteristic as the simple quadratic model which suggests that this methodology and the derived insights can be applied also to advanced models and applied purposes. In total, these examples suggest scenario-dependent interaction patterns between structural and probabilistic ambiguity. We are not aware of such a distinction and presentation in the literature.\medskip 

On a practical note, we suspect that in the modeling work very easily observation scenarios $(i)$ and $(ii)$ are confused, i.e. although having observations of type $(ii)$ the equations of $(i)$ might be used and vice versa. Obviously this leads to wrong judgments based on such defective posteriors and we advise a clear problem understanding. As might have become apparent throughout this work, introducing this clarity about the different observation scenarios facing problem inherent ambiguities is a key motivation and result of this work. Further, although the results for observation scenario $(ii)$ seem disadvantageous compared to $(i)$ these are the right posteriors and the best information level about $\boldsymbol{x}$ given the limited observations (using only one realization of $\boldsymbol{A}$) when correctly incorporating probabilistic ambiguity in the forward model. Lastly, although we present besides the general conceptual derivation direct simulations for observation scenario $(iii)$ future work could expand on this.\medskip

The primary limitation of this methodology in the presented form is computational scalability: Due to the grid-based structure of the input space $\mathbb{R}^n$ and the Monte Carlo integration in the parameter space $\mathbb{R}^K$, large dimensions in either of those may lead to scalability problems as discussed in Subsection \ref{sec:NumSolForm}. To address this, a natural extension would be going from grid-based to problem-specific basis function representations of a high-dimensional input space and smart sampling approaches, such as importance sampling or Markov chain Monte Carlo, e.g. see \cite{Saibaba 2019,Marzouk 2006, Kugler 2022}. We deliberately left this out of the paper since there is a broad literature about variance reduction of Monte Carlo methods and posterior approximation and it is not in the main focus of this more conceptual / methodological presentation. Further questions are: how to deal with situations when the forward model parameter densities are not known well and need to be approximated, and how these extensions play out in practice with higher-dimensional forward models that exhibit a high degree of nonlinear uncertainty propagation? 

\section{Conclusion}

This work shows that highly ambiguous inverse problems arising from the non-injectivity and nonlinearity of the forward model as well as its dependence on multi-modal random variable parameters present a challenging situation. By presenting three substantially different observation scenarios we show through numerical simulations in which situations parts of the ambiguities can be resolved and which parts remain as residual ambiguities. The simulation cases include a family of quadratic models, the inverse parameter estimation of the epidemiological SIR ODE model and the inverse heat source localization of a heat PDE model. The work has a conceptual and explorative character and future research should concentrate on efficient implementations and further developping the ambiguity concept in inverse problems.


\end{document}